\documentclass[10pt]{article}

\usepackage{amsmath}

\begin{document}
\thispagestyle{empty}

\begin{center}
{\bf  A.A.  Bogush, V.M.  Red'kov \\
On unique
parametrization of the linear group $GL(4.C)$ and its\\
 subgroups  by using the Dirac matrix algebra  basis }\\
Institute of Physics, National  Academy of Sciences of Belarus \\
bogush@dragon.bas-net.by, redkov@dragon.bas-net.by
\end{center}

\begin{quotation}

A  unifying overview of the ways to parameterize the linear group
$GL(4.C)$ and its subgroups  is given. As parameters for this
group   there are taken 16 coefficients $G
=G(A,B,A_{k},B_{k},F_{kl})$ in resolving matrix  $G\in GL(4.C)$ in
terms of 16 basic elements of the Dirac matrix algebra. Alternatively to the use of 16 tensor
quantities,  the possibility
to parameterize  the group $GL(4.C)$ with the help of  four 4-dimensional complex vectors
$(k, m,  n, l)$  is investigated.  The
multiplication rules $G'G$ are formulated in the form of a
bilinear function of two sets of 16 variables. The detailed
investigation is restricted to  6-parameter case $G(A,B,F_{kl})$,
which provides  us with spinor  covering for   the complex
orthogonal  group  $SO(3.1.C)$. The complex Euler's  angles
parametrization for the last  group is also given.  Many different
parametrizations of the group based on the curvilinear coordinates
for complex extension of the 3-space of constant curvature are
discussed. The use of the Newmann-Penrose formalism  and applying
quaternion techniques in the theory of complex Lorentz group are
considered. Connections between  Einstein-Mayer  study on
semi-vectors and Fedorov's treatment of the Lorentz group theory
are stated in detail.
Classification of fermions in intrinsic parities is given on the
base of the theory of representations for  spinor covering of the
complex Lorentz group.

\end{quotation}


\section{Introduction}
\hspace{5mm}
Physical applications  especially in connection with
relativity theory and quantum mechanics  extensively employ the
real  orthogonal groups: $SO(3.1.R)$,
 $SO(4.R)$,  and  $SO(2.2.R)$. Also 3-rotation groups
   $SO(3.R)$  and $SO(2.1)$ are widely used.
Much attention was given to the complex orthogonal groups
$SO(4.C)$, including all previous as sub-groups. Surely, from
general mathematical viewpoint, the theory of these orthogonal
groups can be considered to have been successfully  solved  many
years ago and one should  not expect  to obtain new facts. At the
same time, everyday employing of these groups in many different
contexts  requires some elaborate
 and desirably unified apparatus in parameterizing these groups.
In practice, all calculation with those groups involve some
parametrization  of them. It is the more so when calculation is
done  with  finite transformations of the groups. Special and
accurate parametrization for any orthogonal group becomes a point
of first significance when turning to the problem of its covering
spinor groups.

Impressive  progress  had been reached in the original treatment
of the theory of the Lorentz group given by Fedorov F.I. (1979).
Though the author (with co-authors)
 had given a self-closed theory of the  real  Lorentz group and extension of the methods
 used to deal with other orthogonal groups, this treatment stays
 some distance away from  the standard and
 widely used  ways of treatment of the Lorentz group.

In Fedorov approach,  an arbitrary finite Lorentz transformation
in 4-vector space from the very beginning is constructed as a
product of two mutually commuting matrices, each of those is a
linear function of a complex 3-vector (parameter). Then it is
demonstrated that in this way one can obtain all proper and
orthochronous Lorentz transformations in 4-vector space. Six
independent parameters are obtained as a result of imposing
additional restrictions on two complex vectors, involving complex
conjugation. Such approach had been extended to the real groups
$SO(4.R),SO(2.2)$ and complex orthogonal group $SO(3.1.C) \sim
SO(4.C)$. The use of this vector-based parametrization of the
Lorentz group has permitted the solving in full details many of
particular problems where  substantial role is given to
relativistic invariance. The use of the fixed parameters
associated with relativistic transformations, including
arbitrarily  oriented Euclidean 3-rotation and any Lorentz boost,
enables the  possibility  of overcoming  many technical difficulties and
bring calculations to final analytical results in closed
mathematical form.

%

We will now discuss our list of references as they apply to this paper.
The groups under
consideration are now part of everyday physics practice. So a
list of references will be  quite long. Since a discussion of the references
in relation to each other is a rather  daunting
task, we  have chosen an historical principle
and  use classifying by date. In the present paper we give only
several guideline comments and some indications on the content of
only part of works from the reference list, the most directly
referring to subject under consideration\footnote{With the hope
 that a comprehensive analytical review of the  bibliography given will
 be given by us elsewhere.  As a compromise, for conscience  sake, we gave the full
  titles of the papers  cited.}

 After establishing the fundamental role of the Lorentz group in  physics:
Lorentz (1904), Poincare (1904, 1905), Einstein (1905), and after
Minkowski  (1908-1909)
 elaboration of  4-dimensional space-time geometry, the relativistic  4-tensor apparatus
had  been entered into usage in physics. Because the main
relativistic object was the electromagnetic field, the most of the
work was given to this field: Silberstein (1907); Minkowski
(1908-1909) Sommerfeld (1910).

Usually,  invention of spinors  as mathematical objects is given
to Cartan (1913); however it should be noted that  before Cartan
other French   mathematician Darboux had  examined
(1900,1004,1905, 1914)  particular geometric  properties of  the
objects  that in essence represent spinors. These  mathematical
achievements in spinors  had been ignored by physics.
Mathematicians became interested in the theory of continuous
transformation groups: Weyl (1924); Schreier (1925-1926).

There arises the  physical concept of the spin: Uhlenbeck and
Goudsmit (1925); Thomas (1926); Frenkel (1926); Pauli (1927).  In
the context of quantum mechanic, Dirac (1929) had invented a
relativistic wave equation for a particle with spin half-one. In
this connection the interest in 2-spinors and 4-spinors had been
grown much: M\"{o}glich (1928); Neuman (1929); Van der  Waerden
(1929); Laporte and  Uhlenbeck (1931); Rummer (1931). Weyl (1931)
and Wigner (1931) had written their great books on applying the
group theory in quantum physics.
 In the context of Schr\"{o}dinger equation,  Wigner (1932) had
proposed to use the operation of inverting the time
  which included the complex conjugation.
Einstein and Mayer  (1932-1933) gave the original mathematical  treatment for Lorentz
group and its simplest representations, introduced  the concept of semi-vector,  the
4-dimentional objects closely connected with 2-spinors.

 Steady interest to spinors  retained: Mie (1933); Veblen (1933);  Pauli (1933);
 Brauer and Weyl (1935); Sommerfeld ( 1936);  Rumer (1936). Majorana had written  his great
 paper on neutral 4-spinors (1937). This problem gave rise to a new interest to the role
 of complex imaginary unit in quantum theory:  Dirac (1937); Racah (1937); Kramers (1937);
 Hettner (1938); Scherzer (1938); Furay (1938). There appeared the reviewing
  physical and mathematical works:   Cartan (1938); Dirac (1939); Brillouin (1938);  Wigner (1939);
   Pauli and Belinfante (1940); Pauli (1941) Einstein and Bargmann (1944,1947)
    had written two papers on
   possible use of bivectors in the context of a unified theory.

   After evident stopping investigation, the role of relativistic invariance in the
   quantum  theory
   is again of special interest: Gel'fand and Yaglom (1948); Wigner (1948).
   Yang and Tiomno (1950) had written their  paper in which firstly the problem of classifying
   fermions in intrinsic parities had been possed. In the same time there apperaed
   many  others works  studying the same problem:  Zharkov (1950); Wick, Wigner, Wightmann
   (1952);  Pais and Jost (1952);  Shapiro  (1952,1954). Interest to the Lorentz group
   theory and discrete  symmetries is steady:  Bade and Jehle (1953); L\"{u}ders (1954);  Umezava, Kamefuchi,
   Tanaka ( 1954); Naimark (1954); Rashevski (1955);  Good (1955); Case (1955);  Pauli (1955);
    Watanabe (1955);   Gel'fand , Minlos,  Shapiro (1956);
    Umezava (1956).

Non-conservation of parity:  Lee and Yang (1956,1957); Lee,Oehme, Yang (1957); Landau (1957)
 gave an impulse for  many  new  works:
 Golfand (1956); Wigner (1957); Sokolik (1957); Case (1957);  Heine (1957);
  Schremp (1957);   Feinberg and
 Weinberg  (1959).

Practically in the same time, and independently from each other,
there appeared general surveys on the Lorentz group invariance:
Corson (1953);  Winogradzki  (1957-1959);   Van Winter ( 1957);
Shirokov (1957-1959); Naimark (1958); Grawerts , L\"{u}ders,
Rollnik (1959), Halbwachs, Hillion, Vigier (1959); Wightmann
(1960),  Jost (1960,1962);  Fedorov and Bogush (1858, 1961, 1962);
Macfarlane (1962); Wigner (1962,1964).  We can see  several  tendencies
in the manner of working with the Lorentz  group invariance: to
analyze the relativistic invariance in the frames   of  the
corresponding Lie algebra (formalism of infinitesimal
transformations); to parameterize the finite Lorentz group
transformations   with the help  of: 1)  three Euler's complex
angles; 2)  bi-vector; 3) a 4-vector on complex sphere;  4) a complex
3-dimensional vector. All these techniques are in use  up to
present time\footnote{ Intensive  work on constructing the theory
of the Lorentz group and related to it has been done by the
Fedorov's scientific school:
 Fedorov (1967); Fedorov and Tkharev (1968); Ivanitskaya (1969); Fedorov (1970);
 Bogush, Moroz, Fedorov (1970); Bogush and Fedorov (1972); Bogush (1972, 1973); Fedorov (1973);
 Berezin and Fedorov (1975); Fedorovych and Fedorov (1975); Bogush, Kurochkin , Fedorov (1977);
  Bogush,  Fedorov (1977);
 Bogus,  Ivanitskaya, Kisel, Kostjukovich (1978); Berezin, Tolkachev, Fedorov (1970); Berezin ,
 Fedorov (1983); Fedorov (1982). There were written three monographes:
 Fedorov (1979); Ivanitskaya (1979); Berezin, Kurochkin, Tolkachev (1979).}

The task of classifying particles, including fermions, in
intrinsic parities takes attention: Sudarshan (1965); Lee
and Wick (1966);  Goldberg (1969); Harley (1974); Biritz (1975);
Staruszkiewich  (1976); Ebner (1977); Altmann (1986); Sharma
(1989); Gaioli and Alvarez (1995); Kauffman (1996); Half (1996),
Ackermann (1996);  Brumby, Foot, Volkas (1996); Annandan (1998);
Erdem (1998);  Erber (2004). The same can be noted for many other
aspects of the Lorentz group theory; Penrose (1974); Good (1995);
Buchbinder, Gitman, Shelepin (2001); much attention is given the
Majorana object (for instance see: Huet and Neuberger (1996,1998);
Dvoeglazov (1977); on applying of quaternions  see: Stefano,
Rodrigues (1998).

Several words must be added in connection with general spinor
approach in physics and the  idea  on spinor structure of
space-time: Newman and Penrose  (1962); Frolov (1977); Alexeev and
Khlebnikov (1978);
 Penrose and Rindler (1984,1986) - in the last books
  see comprehensive list of references  on the subject.

It is well-known that the space-time vector  $x^{a}$ = $( t, x, y,z)$
can be identified with the explicit realization of the one-valued simplest
representation of the (restricted) Lorentz group
$L^{\uparrow}_{+}$. But in nature we face particles of integer and half-integer spin,
 bosons and fermions. Among respective sets of Lorentz group representations, entities
 describing bosons and fermions,  there exists a clear-cut distinction: boson-based representations
$T_{bos.}$  are single-valued whereas fermion-based representations $T_{ferm.}$  are double-valued
on the group  $L^{\uparrow}_{+}$. In other words, $T_{bos.}$ are global representations, whereas
 $T_{ferm.}$ are just local ones.
 One might have advanced a number of theoretical arguments to neglect  such a slight trouble.
 However, the fact of prime importance is that this global-local difficulty cannot
 be cleared up --
  in the frames  of the orthogonal group $L^{\uparrow}_{+} = SO(3.1)$ it is
     insurmountable.
 It has long been known that to work against the global-local problem one must investigate and employ
  one-valued representations (boson-based as well as fermion-based) of the covering group $SL(2.C)$.
 The group $SL(2.C)$ is other group, different from
$L^{\uparrow}_{+} = SO(3.1)$, but it is linked to  the latter by a
quite definite homomorphic
 mapping. At this every local representation of the orthogonal group has its counterpart --
  global representation of the covering group; evidently,  it is a quite usual procedure.

In essence,  a strong form of this changing $L^{\uparrow}_{+} =
SO(3.1) \;\; \Longrightarrow \;\; SL(2.C)$, when
 instead of the
orthogonal Lorentz group  $L^{\uparrow}_{+} = SO(3.1)$   we are
going to employ the  covering group $SL(2.C)$ and its
representations throughout, and in addition we are going to work
in the same manner  at describing
 the space-time structure itself, is the known Penrose-Rindler spinor  approach.

The idea on spinor structure  of space-time can be of crucial
significance  in our attempts to solve the problem of classifying
the fermions in intrinsic parities. Indeed,  it is natural
requirement: if  one  adopts the space-time with spinor structure,
then  one must consider  the problem of classifying particles in
intrinsic parities in terms of exact linear representations of the
covering for the full Lorentz group $L ^{\uparrow \downarrow}
_{+-}$, including $P$ ant $T$- reflections. This problem was
considered previously by one of authors: Red'kov (1996, 2002). Now
we extend the results to spinor covering for complex Lorentz
group.

The main goal of the present paper is to  give  a brief account of
the ways to use some parameters while working with  the Lorentz
group and the groups related to it. In particular, we are
interested in extending all the apparatus on the complex Lorentz
group  -- on group theoretical ground for  twistors idea
see   Penrose and Rindler (1986).
And finally one other idea was  of value while  preparing this paper: we have presented
one special way  to treat  the theory of  general linear complex
 group $GL(4.C)$ in the line given by the Dirac matrix algebra. On one hand,  we
 obtain extension of the methods used and developed  for Lorentz group theory;
 on  the other hand,  this linear  group includes unitary sub-groups, $SU(4)$
   and related to it, which may be  interesting  in many  physical applications.

\vspace{2mm}

The paper  is organized as follows:

\vspace{2mm}

1. Introduction

2. On linear (tensor and vector)  parametrization of the group  $GL(4.C)$

3. Parametrization of the  complex Lorentz  group  and its
subgroups

4. Complex Euler's angles, coordinates on the  sphere

5. On connection between  Fedorov's construction for  the Lorentz
group  theory

and Einstein-Mayer concept of semi-vectors

6. Complex Lorentz group and quaternions

7. On the use of   Newman-Penrose  formalism

8. The  covering for  $SO(3.1.C)$  and intrinsic fermion parity

9. On the structure of Majorana bases

10. Conclusions

\section{On linear (tensor and vector)  parametrization of the group  $GL(4.C)$}

\hspace{5mm}
We  have started from simple idea: any 4-matrix can be
resolved in terms of 16 basic elements of Dirac matrix algebra, so we represent a matrix
$G \in GL(4.C)$ as linear combination of Dirac ones
\begin{eqnarray}
G = A \; I + iB \; \gamma^{5} + iA_{l}\; \gamma^{l} + B_{l}\;
\gamma^{l} \gamma^{5} + F_{mn} \; \sigma^{mn} \; . \label{1}
\end{eqnarray}

\noindent We can determine a  natural way to parameterize the
linear group $GL(4.C)$  in terms of 16  basic elements of independent complex (tensor) quantities
$G=G(A,B,A_{k},B_{k},F_{kl})$. With the use of
algebraic properties of the Dirac matrices   we have derived the
composition rule  for the introduced parameters of the group  $GL(4.C)$
\begin{eqnarray}
A''= A'\; A -B'\; B -A'_{l} \; A^{l} -B'_{l} \; B^{l} - {1\over 2
} \;F'_{kl} \;F^{kl} \; , \nonumber
\\
B''= A'\; B + B'\; A +  A'_{l} \; B^{l} - B'_{l} \; A^{l} +
{1\over 4 } \;F'_{mn} \;F_{cd}\;
 \epsilon^{mncd}\; ,
\nonumber
\\
A''_{l} = A' \;A_{l} - B'\;B_{l} + A'_{l\;}A + B'_{l}\;B  +
A'^{k}F_{kl} + \nonumber
\\
+ F'_{lk}A^{k} + {1\over 2} \; B'_{k} \;F_{mn} \;
\epsilon_{l}^{\;\;\;kmn} + {1\over 2}\; F'_{mn}\;B_{k} \;
\epsilon_{l}^{\;\;\;mnk} \; , \nonumber
\\
B''_{l} = A'\;B_{l} + B'\;A_{l} - A'_{l}\;B + B'_{l} \;A +
B'^{k}\;F_{kl} + \nonumber
\\
+  F'_{lk} \;B^{k} + {1\over 2} \; A'_{k} \;F_{mn}\;
\epsilon^{kmn}_{\;\;\;\;\;\;\;l} + {1\over 2} \; F'_{mn} \;A_{k}
\; \epsilon^{mnk}_{\;\;\;\;\;\;\;l} \; , \nonumber
\\
F''_{mn} = A'\;F_{mn} + F'_{mn} \;A  - (A'_{m} \; A_{n} -A'_{n} \;
A_{m}) - (B'_{m}\;B_{n} -B'_{n}\;B_{m})  + \nonumber
\\
+ A'_{l}\;B_{k} \; \epsilon^{lkmn} - B'_{l}\; A_{k} \;
\epsilon^{lkmn}\; +{1\over 2} \;  B'\;F_{kl} \;
\epsilon^{kl}_{\;\;\;\;\;mn} + {1\over 2} \; F'_{kl}\;B\;
\epsilon^{kl}_{\;\;\;\;\;mn}  + \nonumber
\\
+ (F'_{mk} \;F^{k}_{\;\;n} -  F'_{nk} \;F^{k}_{\;\;m}  )\; .
\label{2}
\end{eqnarray}

For the following, it is convenient to use (3+1)-splitting:
\begin{eqnarray}
G = A \; I + iB \; \gamma^{5} + iA_{0}\; \gamma^{0} +  iA_{i}\; \gamma^{i}+
\nonumber
\\
 + B_{0}\; \gamma^{0} \gamma^{5} + B_{i}\; \gamma^{i} \gamma^{5}+
 a_{i} \;    K^{i} +  b_{i} \;  J^{i}
\label{2'}
\end{eqnarray}

\noindent where the notation is used
\begin{eqnarray}
{\bf a} = (a_{i}) = (F_{0i}) \; , \qquad
{\bf b} = (b_{i}) = ({1 \over 2} \epsilon _{ikl} F_{kl}) \;  ,
\nonumber
\\
{\bf K} = ( K^{i} )=  (2\sigma^{0i})  \;, \qquad {\bf J} = (J^{i}) =
(\epsilon _{ikl} \sigma^{kl} ) \; .
\nonumber
\end{eqnarray}

\noindent
In spinor basis of Dirac matrices [113]
\begin{eqnarray}
\gamma^{a} = \left | \begin{array}{cc} 0  & \bar{\sigma}^{a}  \\
\sigma^{a} & 0
\end{array} \right | \; , \;\; \sigma^{a}= (I , \sigma^{j}), \bar{\sigma}^{a}= (I , -\sigma^{j}),\;
\gamma^{5}= \left |  \begin{array}{cc} -I & 0 \\ 0 & +I
\end{array} \right | \; ,
\nonumber
\end{eqnarray}

\noindent the matrix  (\ref{2'}) will take the form
\begin{eqnarray}
G   =   \left | \begin{array}{rr}
k_{0}  + \; {\bf k} \; \vec{\sigma}  \;\;&  n_{0}  - \; {\bf n}  \; \vec{\sigma} \\[3mm]
- l_{0}  - \; {\bf l} \; \vec{\sigma} \;\;  & m_{0} - \; {\bf m}
\; \vec{\sigma}
\end{array} \right | \; ,
\label{6.2}
\end{eqnarray}

\noindent
where the notation is used:
\begin{eqnarray}
k_{0} = a_{0} - i b_{0} \;, \qquad  {\bf k} = {\bf a} - i {\bf b} \;,
\nonumber
\\
m_{0} = a_{0} + i b_{0} \;, \qquad  {\bf m} = {\bf a} + i {\bf b} \; ,
\nonumber
\\
 l_{0} =B_{0} - i A_{0}  , \qquad    {\bf l} = {\bf B} - i {\bf A}  \; ,
\nonumber
\\
n_{0} = B_{0} + i A_{0}   , \qquad  {\bf n}   = {\bf B} + i {\bf A}  \;
.
\label{2.9b}
\end{eqnarray}

\noindent
In these parameters $(k,m;n,l)$ the composition rule (\ref{2})
\begin{eqnarray}
(k'',m'';n'',l'') =< (k',m';n',l'), (k,m;n,l) > \; ;
\nonumber
\end{eqnarray}

\noindent
will take the form
\begin{eqnarray}
k_{0}'' = k_{0}' \; k_{0} + {\bf k}' \; {\bf k}
 - n'_{0}\;  l_{0} + {\bf n}' \; {\bf l} \;,
\nonumber
\\
{\bf k}'' = k'_{0} \; {\bf k} + {\bf k}' \; k_{0}  + i \;  {\bf
k}' \times {\bf k} - n_{0}' \; {\bf l} + {\bf n}'\; l_{0} + i\;
{\bf n}' \times {\bf l} \; ,
\nonumber
\\
m_{0}'' = m_{0}' \;  m_{0} + {\bf m}' \; {\bf m}
 - l'_{0}\;  n_{0} + {\bf l}' \; {\bf n} \;,
\nonumber
\\
{\bf m}'' = m'_{0} \; {\bf m} + {\bf m}' \; m_{0}  - i \;  {\bf
m}' \times {\bf m} - l_{0}' \; {\bf n} + {\bf l}'\; n_{0} - i\;
{\bf l}' \times {\bf n} \; ,
\nonumber
\\
n_{0}'' = k_{0}' \; n_{0} - {\bf k}' \; {\bf n}
 + n'_{0} \; m_{0} + {\bf n}' \; {\bf m} \;,
\nonumber
\\
{\bf n}'' = k'_{0} \; {\bf n} - {\bf k}' \; n_{0}  + i \;  {\bf
k}' \times {\bf n} + n_{0}' \; {\bf m} + {\bf n}'\; m_{0} - i\;
{\bf n}' \times {\bf m} \; ,
\nonumber
\\
l_{0}'' = l_{0}' \;  k_{0} + {\bf l}' \; {\bf k}
 + m'_{0}  \; l_{0} - {\bf m}' \; {\bf l} \;,
\nonumber
\\
{\bf l}'' = l'_{0} \; {\bf k} + {\bf l}' \; k_{0}  + i \;  {\bf
l}' \times {\bf k} + m_{0}' \; {\bf l} - {\bf m}'\; l_{0} - i\;
{\bf m}' \times {\bf l } \; ;
\label{2.13}
\end{eqnarray}

\vspace{5mm}

The formulas (\ref{2.13})  are true for any matrices from the group  $GL(4.C)$.
This parametrization could be the base for theory of the special linear  groups
$SL(4.C)$ and $SL(4.R)$,  and also for the theory of  the unitary groups $SU(4),
SU(3.1),SU(2.2)$; they   will be
considered separately elsewhere.

\section{Parametrization of the  complex Lorentz  group  and its
subgroups}

Below  we will restrict ourselves to the case of $4\times 4$ matrices when
$$
G(A,B,A_{k}=0,B_{k}=0,F_{kl}) , \qquad or \qquad G(k,m;0,0) \; ,
$$

\noindent i.e. consider  the 8-parametric $4\times 4$ matrices in the quasi diagonal form
\begin{eqnarray}
G   =   \left | \begin{array}{cc}
k_{0}  + \; {\bf k} \; \vec{\sigma}  \;\;&  0 \\[3mm]
0 & m_{0} - \; {\bf m}
\; \vec{\sigma}
\end{array} \right | \; .
\label{6.2'}
\end{eqnarray}

\noindent The composition rules  (\ref{2})    and  (\ref{2.13})  are reduced respectively to
 \begin{eqnarray}
A''= A'\; A -B'\; B  - {1\over 2 } \;F'_{kl} \;F^{kl} \; ,
\nonumber
\\
B''= A'\; B + B'\; A + {1\over 4 } \;F'_{mn}\; F_{cd}
\;\epsilon^{mncd}\; , \nonumber
\\
F''_{mn} =  (\; A'\;F_{mn} + F'_{mn}\; A \; )+
 (\; {1\over 2} \;  B'\;F_{kl} \; \epsilon^{kl}_{\;\;\;\;\;mn} +
 \nonumber
 \\
 +
{1\over 2} \; F'_{kl}\;B\; \epsilon^{kl}_{\;\;\;\;\;mn} \; )  +
(F'_{mk} \;F^{k}_{\;\;n} -  F'_{nk}\; F^{k}_{\;\;m}  )\; ;
\label{3}
\end{eqnarray}

\noindent and
\begin{eqnarray}
k_{0}'' = k_{0}' \;k_{0} + {\bf k}' \;{\bf k} \; , \qquad\;\; {\bf
k} '' = k_{0}' \; {\bf k} + {\bf k}' \;  k_{0} + i \; {\bf k}'
\times {\bf k} \; ;
\nonumber
\\
m_{0}'' = m_{0}' \;m_{0} + {\bf m}' \;{\bf m} \; , \qquad {\bf m}
'' = m_{0}' \; {\bf m} + {\bf m}' \;  m_{0} - i \; {\bf m}' \times
{\bf m} \; .
\label{3.6b}
\end{eqnarray}

\noindent With    two additional constraints  on  8 quantities
$(A,B,F_{kl})$:
 \begin{eqnarray}
 k_{0}^{2} - {\bf k}^{2} = +1 , \qquad   m_{0}^{2} - {\bf m}^{2} = +1 \; ,
 \qquad \qquad  \mbox{or}
\label{4}
\\
 A\; B  -
{\bf a}\; {\bf b}    =0  \; , \qquad ( A^{2} -  B^{2})  -   ({\bf
a}^{\;2} - {\bf b}^{\;2} ) =+1 \; \nonumber
\end{eqnarray}

\noindent
 we will arrive at a definite way to parameterize
spinor covering for complex Lorentz group  $SO(3.1.C) \sim SO(4.C)$,
treated here as a sub-group of the linear
 group $GL(4.C)$ .
At this, the problem of inverting the $G$ matrices is easily solved: in
tensor  and vector forms:
\begin{eqnarray}
G = G(A,B,F_{kl}) \; , \qquad  G^{-1} = G(A,B,-F_{kl}) \; ,
\nonumber
\\
G=G(k_{0}, {\bf k}, m_{0},{\bf m}) , \qquad G^{-1}=G(k_{0}, -{\bf
k}, m_{0},-{\bf m}). \label{6}
\end{eqnarray}

\noindent
Transition from covering 4-spinor transformations to 4-vector ones
is  performed through the  known relationship
\begin{eqnarray}
G\gamma^{a}G^{-1} =
\gamma^{c}L_{c}^{\;\;a}
\nonumber
\end{eqnarray}

\noindent
which  determine $2 \Longrightarrow 1$
map from $\pm G$ to $L$. At this an explicit form of the Lorentz
vector matrices is derived, they are bilinear functions of
$(A,B,F_{kl})$ or $(k_{a},m_{a})$:
\begin{eqnarray}
L_{0} ^{\;\;0} = k_{0} \; m_{0} +  k_{j} \; m_{j} \; , \nonumber
\\
(L_{j}^{\;\; 0} ) = -k_{0} \;  m_{j} -m_{0} \; k_{j} - i\;
\epsilon_{jln} \; k_{l} \;m_{n}  \; , \nonumber
\\
(L_{0}^{\;\; j}) = -k_{0} \;  m_{j} -m_{0} \;  k_{j} + i\;
\epsilon_{jln} \;  k_{l} \;
 m_{n}\; ,
\nonumber
\\
L_{l}^{\;\;j} = (k_{0} m_{0} -k_{n} m_{n}) \; \delta_{lj} +
\nonumber
\\
+ (m_{l} \;  k_{j} + k_{l} \; m_{j}) +
 i \; \epsilon_{ljn} \;  ( k_{0}  \; m_{n} -  m_{0} \; k_{n}  )\; .
\label{Lorentz}
\end{eqnarray}

There exist the direct connection between the above 4-dimensional vector parametrization
of the spinor group $G (k_{a},m_{a})$ and
 the Fedorov's parametrization [100, 137]  of the  group of  complex orthogonal Lorentz transformations
 with the help of 3-dimensional vectors (to the case of real Lorentz group corresponds
 imposing an  additional constrain  $m_{a} =k_{a}^{*}$ or ${\bf M} = {\bf Q}^{*}$):
\begin{eqnarray}
{\bf Q} = { \pm {\bf  k} \over  \pm k_{0}} \; , \qquad {\bf M} = {
\pm {\bf  m} \over \pm m_{0}}\;, \qquad
k_{0} =  \sqrt{ 1 + {\bf k}^{2}}\; , \qquad   m_{0} =  \sqrt{ 1 +
{\bf m}^{2}}\;; \label{7}
\end{eqnarray}

\noindent with the  simple  composition rules for  vector parameters\footnote{Slight  difference
with formulas from  [137] can be eliminated
by  the formal change ${\bf Q} \Longrightarrow i {\bf Q}, \; {\bf M} \Longrightarrow i {\bf M}$.}
\begin{eqnarray}
 {\bf  Q} '' = {  {\bf Q} + {\bf Q}'  + i \; {\bf Q}'
\times {\bf Q}  \over 1  + {\bf Q}' \;{\bf Q}}
\; ;\qquad
 {\bf M} '' = { {\bf M} + {\bf M}'  - i \; {\bf M}' \times
{\bf M} \over 1 + {\bf M}' \;{\bf M} }\; ;
\end{eqnarray}

Evidently,  this pair $({\bf Q}, {\bf M})$ provides us with
possibility to parameterize correctly orthogonal matrices only.
Instead, the $(A,B,F_{kl})$ or $(k_{a},m_{a})$ represent  correct
parameters for the spinor covering group.

It should be noted that there exists one other way to parameterize
4-spinor (covering)
 matrices by means of two complex vectors without any additional restrictions on them -- it was
 given in [139]:
 \begin{eqnarray}
{\bf  p} = {{\bf k} \over 1 + k_{0}} \; , \qquad {\bf  s} = {{\bf
m} \over 1 + m_{0}} \; . \label{8}
\end{eqnarray}

\noindent
It is readily verified that the sign distinction in pure
spinor matrices transforms into the form
\begin{eqnarray}
(k_{0}, {\bf k}) \; \Longrightarrow \; {\bf p} , \qquad (-k_{0},
-{\bf k}) \; \Longrightarrow \; {{\bf p} \over {\bf p}^{2}}  ={\bf
p}' \; , \nonumber
\\
(m_{0}, {\bf m}) \; \Longrightarrow \; {\bf s} , \qquad (-m_{0},
-{\bf m}) \; \Longrightarrow \; {{\bf s} \over {\bf s}^{2}}  ={\bf
s}' \; . \label{10}
\end{eqnarray}

 Sometimes, when we are
interested only in local properties of the spinor representations
then no substantial differences between orthogonal groups and
their spinor  coverings exist. However, in opposite cases global
difference between orthogonal and spinor groups may be very
substantial as well as correct parametrization of them.

\vspace{3mm}

One may compile the table in which all parametrizations used  are
listed:

\vspace{3mm}

\underline{ complex tensors $A,B,F_{kl}$}

   (two subsidiary conditions   are imposed; spinor and orthogonal
   groups are parameterized);

\vspace{3mm}

\underline{ complex bi-vector $f_{kl}$}

   (no  subsidiary conditions are impose; orthogonal group only
   are parameterized);

\vspace{3mm}

\underline{two  complex 4-vectors $k_{a}$ and $m_{a}$}

   (two subsidiary conditions are imposed;  spinor and orthogonal
   group are parameterized);

\vspace{3mm}

\underline{two  complex 3-vectors ${\bf Q}$ and ${\bf M}$}

   (no subsidiary conditions imposed; orthogonal group only is
   parameterized);

\vspace{3mm}

\underline{two  complex 3-vector ${\bf p}$ and ${\bf s}$}

   (no subsidiary condition are imposed; spinor and orthogonal
   group are parameterized);

\vspace{3mm}

\underline{complex Euler angles} $(\alpha, \beta, \gamma )$

    (curvilinear coordinates with no subsidiary conditions,

 different domains for spinor and orthogonal matrices);

\vspace{3mm}

\underline{complex curvilinear coordinates  $(u^{1},u^{2},u^{3})$
on   $SO(4.C)$ }

(34 types of curvilinear coordinates,

 no subsidiary conditions,

 different domains for spinor and orthogonal matrices);

\vspace{5mm} Restrictions specifying the  spinor coverings for
 orthogonal sub-groups $$ SO(3.1.R), SO(2.2.R), SO(4.R),
SO(3.C),  SO(3.R), SO(2.1.R)$$ are well known [118,123,. In
particular, restriction to \underline{real} Lorentz group
$O(3.1,R)$  is achieved by imposing one condition
 (including complex conjugation)
\begin{eqnarray}
(k,m) \qquad \Longrightarrow \qquad (k, k^{*}) \; .
\label{14}
\end{eqnarray}

\noindent The case of    \underline{real}  orthogonal group
$O(4.R)$ is achieved by a formal change ( transition to real
parameters)
\begin{eqnarray}
(k_{0}, {\bf k}) \; \Longrightarrow \; (k_{0}, i\;{\bf k}) \; ,
\qquad (m_{0}, {\bf m}) \; \Longrightarrow \; (m_{0}, i\;{\bf m})
\; , \label{15}
\end{eqnarray}

\noindent and the   \underline{real}  orthogonal group $O(2.2,R)$
corresponds transition to real parameters according to
\begin{eqnarray}
(k_{0},  k_{1}, k_{2}, k_{3} ) \qquad \Longrightarrow \qquad
(k_{0},  k_{1}, k_{2}, ik_{3} ) \; , \nonumber
\\
(m_{0},  m_{1}, m_{2}, m_{3} ) \qquad \Longrightarrow \qquad
 (m_{0},  m_{1}, m_{2}, im_{3} ) \; .
\label{16}
\end{eqnarray}

\noindent
  Expressions  for vector and spinor (covering) matrices concurrently
for all these groups can be readily written  in explicit form.

\section{Complex Euler's angles, coordinates on the  sphere}

\hspace{5mm} To parameterize 4-spinor and 4-vector transformations
of the complex Lorentz
 group  one may use curvilinear coordinates.
The simplest and widely used ones are Euler's complex angles
[64,  70, 90-93, 137, 144].
 This possibility is closely connected with  cylindrical coordinates on the complex
 3-sphere \footnote{In general, on the base of the analysis given by Olewski
 [162] on coordinates in the
 Lobachevski space, one can
 propose 34 different complex  coordinate systems appropriate to
 parameterize the complex Lorentz group and its 4-spinor
 covering.}.
Such complex cylindrical coordinates can be  introduced by the
following relations:
\begin{eqnarray}
k_{0} = \cos \rho \cos z \; , \; k_{3} =  i \cos \rho \sin z \; ,
\qquad k_{1} =  i \sin \rho \cos \phi \; , \; k _{2} = i \sin \rho
\sin \phi \; , \nonumber
\\
m_{0} = \cos R \cos Z \; , \; m_{3} =  i \cos R \sin Z \; , \qquad
m_{1} =  i \sin R \Phi \; , \; m _{2} = i \sin R \sin \Phi \; .
\label{11}
\end{eqnarray}

\noindent Here 6 complex variables are independent, $ (\rho,
z,\phi) \; , \; (R, Z ,\Phi) $, additional restrictions are
satisfied identically by definition. Instead of cylindrical
coordinates т  $(\rho, z, \phi)$  и $(R, Z ,\Phi)$ one can
introduce Euler's complex variables $(\alpha,\beta,\gamma)$ and
$(A,B,\Gamma)$ through the simple linear formulas:
\begin{eqnarray}
\alpha = \phi + z \; , \qquad \beta = 2 \rho \; , \qquad \gamma =
\phi - z \; , \nonumber
\\
A  = \Phi + Z \; , \qquad B  = 2 R \; , \qquad \Gamma = \Phi - Z
\; . \label{12}
\end{eqnarray}

\noindent Euler's angles  $(\alpha, \beta, \gamma)$  and
$(A,B,\Gamma)$ are referred to   $k_{a},m_{a}$-parameters  by the
formulas
\begin{eqnarray}
\cos \beta = k_{0}^{2} - k_{3}^{2} + k_{1}^{2} + k_{2}^{2} \; ,
\qquad \sin \beta = 2 \; \sqrt{k_{0}^{2} - k_{3}^{2}} \;
\sqrt{-k_{1}^{2} -k_{2}^{2}} \; , \nonumber
\\
\cos \alpha = { -i k_{0} k_{1} + k_{2} k_{3} \over \sqrt{k_{0}^{2}
- k_{3}^{2}} \; \sqrt{-k_{1}^{2} - k_{2}^{2}} } \; , \qquad \sin
\alpha = { -ik_{0} k_{2} - k_{1} k_{3} \over \sqrt{k_{0}^{2} -
k_{3}^{2}} \; \sqrt{-k_{1}^{2} -k_{2}^{2}} } \; , \nonumber
\\
\cos \gamma = { -i k_{0} k_{1} - k_{2} k_{3} \over \sqrt{k_{0}^{2}
- k_{3}^{2}} \; \sqrt{-k_{1}^{2} -  k_{2}^{2}} } \; , \qquad \sin
\gamma = { -i k_{0} k_{2} + k_{1} k_{3} \over \sqrt{k_{0}^{2} -
k_{3}^{2}} \; \sqrt{-k_{1}^{2} - k_{2}^{2}} } \; ; \nonumber
\\[3mm]
\cos B = m_{0}^{2} - m_{3}^{2} + m_{1}^{2} + m_{2}^{2} \; , \qquad
\sin B = 2 \; \sqrt{m_{0}^{2} - m_{3}^{2}} \; \sqrt{-m_{1}^{2}
-m_{2}^{2}} \; , \nonumber
\\
\cos A  = { +i m_{0} m_{1} + m_{2} m_{3} \over \sqrt{m_{0}^{2} -
m_{3}^{2}} \; \sqrt{-m_{1}^{2} - m_{2}^{2}} } \; , \qquad \sin A
= { +im_{0} m_{2} - m_{1} m_{3} \over \sqrt{m_{0}^{2} - m_{3}^{2}}
\; \sqrt{-m_{1}^{2} -m_{2}^{2}} } \; , \nonumber
\\
\cos \Gamma = { +i m_{0} m_{1} - m_{2} m_{3} \over \sqrt{m_{0}^{2}
- m_{3}^{2}} \; \sqrt{-m_{1}^{2} -  m_{2}^{2}} } \; , \qquad \sin
\Gamma = { +i m_{0} m_{2} + m_{1} m_{3} \over \sqrt{m_{0}^{2} -
m_{3}^{2}} \; \sqrt{-m_{1}^{2} - m_{2}^{2}} } \; . \label{13}
\end{eqnarray}

Complex Euler's angles as parameters for complex Lorentz group
have distinguished feature: 2-spinor constituents are factorized
into  three elementary Euler's transforms ($\sigma^{i}$ stands for
the known Pauli matrices):
\begin{eqnarray}B(k) = e^{-i\sigma^{3} \alpha/2} e^{i \sigma^{1}\beta/2} e^{+i\sigma^{3} \gamma/2}  \; ,
\qquad
B(\bar{m}) = e^{-i\sigma^{3} \Gamma/2} e^{i \sigma^{1}B/2} e^{+i\sigma^{3} A/2}  \; .
\label{Euler}
\end{eqnarray}

\section{On connection between  Fedorov's construction for  the
Lorentz group  theory and Einstein-Mayer concept of semi-vectors}

 \hspace{5mm}
 The main relations serving the initial points for
systematically constructing the Lorentz group theory in the Fedorov approach  and its
generalization   now can be straightforwardly  written down --
 the origin of
these is evident when turning to the spinor covering matrices
specified in Weyl basis,  where they have a quasi diagonal
structure (\ref{6.2'}).
For the complex  Lorentz group it looks as follows
\begin{eqnarray}
B(k) = \sigma^{a} k_{a} , \; B(m) = \sigma^{a}
m_{a}, \bar{m} = (m_{0}, -m_{i}) \; ,
\nonumber
\\
G(k,m)= \left | \begin{array}{cc} B(k) & 0 \\ 0 & B (\bar{m})
\end{array} \right |  \;  ,
\nonumber
\\
G(k,I) = \left | \begin{array}{cc} B(k) & 0 \\ 0 & I  \end{array}
\right | , \;  G(I,m) = \left | \begin{array}{cc} I  & 0 \\
0 & B(\bar{m})  \end{array} \right | \; , \nonumber
\\
 G(k,I) G(I,m) =   G(I,m)  G(k,I) = G(k,m) \; ,
\nonumber
\\
L(k,I) L(I,m) =   L(I,m)  L(k,I) = L(k,m) \; .
\label{17}
\end{eqnarray}

Explicit forms of   $L(k,I)$ and  $L(I,m)$  are
\begin{eqnarray}
L(k,I)= \left | \begin{array}{rrrr}
k_{0} &  -k_{1} & -k_{2}  &  -k_{3} \\
-k_{1}&  k_{0}  &  -ik_{3}       & ik_{2}\\
-k_{2}&  ik_{3}       & k_{0}   &-ik_{1}\\
-k_{3}&  -ik_{2}       &    ik_{1}     & k_{0}
\end{array} \right | , \qquad
L(k,I)= \left | \begin{array}{rrrr}
m_{0} &  -m_{1} & -m_{2}  &  -m_{3} \\
-m_{1}&  m_{0}  &  im_{3}       & -im_{2}\\
-m_{2}&  -im_{3}       & m_{0}   &im_{1}\\
-m_{3}&  im_{2}       &    -im_{1}     & m_{0}
\end{array} \right | .
\label{18}
\end{eqnarray}

In the case of real Lorentz  group, when
 $m = k^{*}$,  the above relations will take the form
\begin{eqnarray}
 G(k,I) G(I,k^{*}) =   G(I,k^{*})  G(k,I) = G(k,k^{*}) \; ,
\nonumber
\\
L(k,I) L(I,k^{*}) =   L(I,k^{*})  L(k,I) = L(k,k^{*}) \; ,
\nonumber
\\
 L(I,k^{*}) = [L(k,I)]^{*} \; .  \qquad
 \label{19}
\end{eqnarray}

It should especially be noted that  existence of
such a factorized structure in the theory of Lorentz group has long been known:
 it appeared in 1932 when
 Einstein and Mayer had published a paper of  a series of these  [26-28]) on the
 so called semi-vectors, special 4-dimensional complex quantities, 1-st and 2-nd  types,
  transformed by means  of  complex matrices   $L(k,I)$  or   $L(I,k^{*})$ respectively.

\hspace{5mm} Complex 4-dimension quantities transformed  with the
help of the matrices $L(k,I)$ and $L(I,m)$
 by definition are  called semi-vectors of 1-st and 2-nd
type (respectively $U$ and  $V$ semi-vector):
\begin{eqnarray}
U' = L(k,I)  U \; , \qquad V' = L(I,m) \; V \; . \label{16.1b}
\end{eqnarray}

With the help of special linear transformations,
 the matrices defining the transformation laws for
semi-vector can be changed to quasi diagonal form.
Indeed  (see  [137])
\begin{eqnarray}
\left | \begin{array}{c} U_{0} \\ U_{1} \\  U_{2} \\  U_{3}
\end{array} \right | =  {1 \over \sqrt{2}}
 \left | \begin{array}{rrrr}
 1 &  0 & 1 &  0 \\
 0 & -1 & 0 & 1 \\
 0 &  i & 0 & i \\
-1 &  0 & 1 & 0
 \end{array} \right | \;
 \left | \begin{array}{c}
 \xi^{1} \\ \xi^{2} \\ \Xi_{1} \\ \Xi_{2}  \end{array} \right | \; ,
  \qquad U = F \; \Psi \; ,
\nonumber
\\
U ' = L(k,I) U : \qquad \Longrightarrow \qquad \Psi' = (F^{-1} L F
)\; \Psi \; ,
\nonumber
\\
F^{-1} L(k,I) F =  \left | \begin{array}{cc} B(k) & 0 \\ 0 &
\sigma^{2} B(k) \sigma^{2} \end{array} \right | \; .
\label{16.8b}
\end{eqnarray}

\noindent This means that  the semi-vector $U$ is equivalent to a
direct sum of  the following 2-spinors (with differently
positioned indices):
\begin{eqnarray}
U \; \Longrightarrow \;\; (\xi^{1},\xi^{2}) \oplus (\Xi_{1},
\Xi_{2}) \;. \label{16.8c}
\end{eqnarray}

Now analogously one should consider  $V$  semi-vector transformed
by the matrix $L(I,m)$:
\begin{eqnarray}
\left | \begin{array}{c} V_{0} \\ V_{1} \\ V_{2} \\ V_{3}
\end{array} \right | =
 {1 \over \sqrt{2}} \;
\left | \begin{array}{rrrr}
0   & 1  & 0  &  -1 \\
1   & 0  & 1  &  0  \\
i  &  0 &  -i  &  0 \\
0  & -1  &  0 & -1
\end{array} \right | \; \left | \begin{array}{c}
\eta_{\dot{1}} \\
\eta_{\dot{2}}\\
H^{\dot{1}}\\
H^{\dot{2}}
\end{array} \right | \; ,
\nonumber
\\
V = L(I,m) V : \qquad \Longrightarrow \qquad \Phi' = (\varphi^{-1}
L \varphi )\; \Phi\; ,
\nonumber
\\
\varphi^{-1} L(I,m) \varphi =  \left | \begin{array}{cc}
B(\bar{m}) & 0 \\ 0 & \sigma^{2} B(\bar{m}) \sigma^{2} \end{array}
\right | \; . \label{16.15b}
\end{eqnarray}

\noindent In other words, the semi-vector  $V$ is equivalent to a
direct sum of two following spinors:
\begin{eqnarray}
V \; \Longrightarrow \;\; (\eta_{\dot{1}},\eta_{\dot{2}}) \oplus
(H^{\dot{1}}, H^{\dot{2}}) \;. \label{16.16c}
\end{eqnarray}

Restriction to real Lorentz group is achieved by imposing
$m=k^{*}$, therefore the spinor
 $\eta$ transforming with the help of the law
 \begin{eqnarray}
 \eta = \left | \begin{array}{c}
\eta_{\dot{1}} \\  \eta_{\dot{2}}
\end{array} \right | \; , \qquad  \eta '= B^{+} (\bar{k}) \; \eta \; ;
\nonumber
\end{eqnarray}

\noindent which corresponds to a standard convention on spinor
(doted and non-doted) indices [113].

\section{Complex Lorentz group and quaternions }

\hspace{5mm}
 Lorentz matrices (\ref{18}) can be represented in
a short algebraic form in the frame of quaternion notation (see for example in [90-93, 130,
138, 144, 156, 157]):
\begin{eqnarray}
L(k,I) = k_{0} e_{0} + k_{i} e_{i} = k_{a} e_{a}  \; ,
\qquad
L(I,m) = m_{0} e^{*}_{0} + m_{i} e^{*}_{i} = m_{a} e^{*}_{a} \; ,
\label{20}
\end{eqnarray}

\noindent
where (the asterisk symbol  $*$ stands for complex conjugation)
\begin{eqnarray}
e_{0} = \left | \begin{array}{cccc}
1 & 0 & 0 & 0 \\
0 & 1 & 0 & 0 \\
0 & 0 & 1 & 0\\
0 & 0 & 0 & 1
\end{array} \right | \;, \qquad
e_{1} = \left | \begin{array}{cccc}
0 & -1 & 0 & 0 \\
-1 & 0 & 0 & 0 \\
0 & 0 & 0 & -i\\
0 & 0 & i & 0
\end{array} \right | \; ,
\nonumber
\\
e_{2} = \left | \begin{array}{cccc}
0 & 0 & -1 & 0 \\
0 & 0 & 0 & i \\
-1 & 0 & 0 & 0\\
0 & -i & 0 & 0
\end{array} \right | \;, \qquad
e_{3} = \left | \begin{array}{cccc}
0 & 0 & 0 & -1 \\
0 & 0 & -i & 0 \\
0 & i & 0 & 0\\
-1 & 0 & 0 & 0
\end{array} \right | \; .
\label{21}
\end{eqnarray}

\noindent
The quaternion quantities
 $e_{a}$ и $ e_{a}^{*}$ obey the rules:
 \begin{eqnarray}
e_{k} e_{l} = \delta _{kl}  + i \epsilon _{kln} e_{n} ,\qquad
e^{*}_{k} e^{*}_{l} = \delta _{kl}  - i \epsilon _{kln} e^{*}_{n}
, \qquad
e_{i} e_{j}^{*} =  e_{j}^{*} \;  e_{i} \; . \label{22}
\end{eqnarray}

The  complex Lorentz matrix can be resolved in terms of
 (double) quaternion units $e_{a}e^{*}_{b}$:
\begin{eqnarray}
L(k,m)  = (k_{0}  + k_{i} e_{i})(m_{0}  + m_{j} e^{*}_{j} )=
\nonumber
\\
= k_{0} m_{0} +  ( k_{0} m_{j} \; e^{*}_{j} + m_{0} k_{i}  \; e_{i} ) +
k_{i}  m_{j} \; e_{i}  e^{*}_{j} \; ;
\label{14.6}
\end{eqnarray}

\noindent
in case of real Lorentz group, when
 $m_{a}= k_{a}^{*}$, the   previous relation will look
\begin{eqnarray}
L(k,k^{*} )  =  k_{0} k^{*}_{0} +   k_{0} k^{*}_{j} \; e^{*}_{j} +   k_{i} k^{*}_{0}  \; e_{i}  +
k_{i}  k^{*}_{j} \; e_{i}  e^{*}_{j} \; .
\label{14.7a}
\end{eqnarray}


Let us briefly discuss one special application of the quaternions in the Lorentz group theory.
If the vector Lorentz matrix is done explicitly, one should find
a corresponding 4-spinor matrix, that is  two parameters
$(k,k^{*})$  of $G$.
The  above structure of  real matrix   $L(k,k^{*})$ points out another simple way to treat
the same  problem,
 which is based on looking at the coefficients at quaternion combinations
$e_{a}e^{*}_{b}$.

\begin{quotation}
From known
$$
\mid k_{0} \mid ^{2} \; I , \qquad
\mid k_{1} \mid ^{2} e_{1} e_{1}^{*}, \qquad
\mid k_{2} \mid ^{2}e_{2} e_{2}^{*} , \qquad
\mid k_{3} \mid ^{2}e_{3} e_{3}^{*} \; ;
$$

\noindent one can determine the modulus
 $\mid k_{a} \mid$,  and then  from
 $$
(k_{0}k_{1}^{*} ) \; e_{1}^{*} \; , \; (k_{0}^{*} k_{1} ) \; e_{1}, \qquad
(k_{0}k_{2}^{*} ) \; e_{2}^{*} \; ,\; (k_{0}^{*} k_{2} ) \; e_{2}, \qquad
(k_{0}k_{2}^{*} ) \; e_{2}^{*} \; , \;
(k_{0}^{*} k_{2} ) \; e_{3}
$$

\noindent one may  find the  phases of  $k_{a}$.

\end{quotation}

The same  problem may be posed for  the complex Lorentz  group as well.
Starting from a given complex matrix $L(k,m)$ one may wish to find two parameters $k$ and $m$.
Let an algorithm to establish the numerical matrix consisting of the coefficient at
quaternion combinations $e_{a}e_{b}^{*}$) be known:
\begin{eqnarray}M_{ab}= k_{a}m_{b} \; .
\label{14.8}
\end{eqnarray}

\noindent
Having the matrix $M$ be given, one may solve the problem as
follows.

\vspace{5mm}
 1)From elements of first line  (with $\pm$ sign
ambiguity) one finds the $k_{0}$:
\begin{eqnarray}
k_{0} = \pm \; \sqrt{ (k_{0}m_{0})^{2} - (k_{0}m_{1})^{2} -
 ( k_{0}m_{2} )^{2} - ( k_{0}m_{3})^{2} }
\label{14.8a}
\end{eqnarray}

\noindent

2) From elements of the first row  one finds remaining three
components $k_{1},k_{2},k_{3}$:
\begin{eqnarray}
k_{1} = k_{0}  \; {M_{10} \over M_{00}}\; , \qquad k_{2} = k_{0}
\; {M_{20} \over M_{00}}\; , \qquad k_{3} = k_{0}  \; {M_{30}
\over M_{00}}\; . \label{14.8b}
\end{eqnarray}

3) From elements of first row one finds  $m_{0}$:
\begin{eqnarray}
m_{0} = \pm \; \sqrt{ (k_{0}m_{0})^{2} - (k_{1}m_{0})^{2} -
 ( k_{2}m_{0} )^{2} - ( k_{3}m_{0})^{2} } \; , \qquad m_{0} k_{0} = M_{00} \;
\label{14.9a}
\end{eqnarray}

\noindent the sign  ($+$ or $-$ )  of  $m_{0}$   must be consistent
with the sign of  $k_{0}$.

 4)
 From elements of the first line one finds three components $m_{1},m_{2},m_{3}$:
\begin{eqnarray}
m_{1} = m_{0} \;  {M_{01} \over M_{00} }\, \qquad m_{2} = m_{0} \;
{M_{02} \over M_{00}}\; \qquad m_{3} = m_{0} \;  {M_{03} \over
M_{00}} \; . \label{14.9b}
\end{eqnarray}

\vspace{5mm}

The matrix $M$ has a diad-structure (\ref{14.8}) that leads to the equations
(one should take into account the identities:
 $k\bar{k} =1,\;  \bar{k} = \delta \; k, m\bar{m} =1,\;  \bar{m} = \delta \; m$):
\begin{eqnarray}
M_{ab} \bar{m}_{b} =  k_{a} \; , \qquad  M_{cb} \bar{k}_{c} =
m_{b} \label{14.10a}
\end{eqnarray}

From these one can easily reduce the task to determine $k$ and $m$
from  given $L(k,m)$ to a pair of eigen-value problems. Indeed,
eqs. (\ref{14.10a}) can be rewritten  i n matrix form
\begin{eqnarray}
M \; \bar{\delta} \; m = k\; , \qquad \tilde{M} \; \bar{\delta} \;  k =
m \; ,
\nonumber
\end{eqnarray}

\noindent from where
\begin{eqnarray}
(M \bar{\delta}  \; \tilde{M} \bar{\delta}) \;  k   = k\; , \qquad
(\tilde{M} \bar{\delta}  \;   M \bar{\delta})\;  m   = m \; .
\label{14.11}
\end{eqnarray}

\noindent Thus, we have constructed the matrices for which
the 4-vector parameters  $k$ and  $m$ are eigen-vectors.

\section{On the use of   Newman-Penrose  formalism }

\hspace{5mm} In the real Lorentz group theory,  especially in
connection to the problems particle  fields in general relativity
the wide use is found  the so-called isotropic basis (also commonly
known
as Newman-Penrose formalism of the light tetrad [163-168]). This
basis can be effectively employed in the the theory of complex
Lorentz  group as well.
Let us turn again to the factorized form (\ref{17}) of the complex Lorentz
matrices
\begin{eqnarray}
L(k,m) =  L(k,I)L(I,m) = L(I,m)L(k,I) ,
\nonumber
\end{eqnarray}

\noindent and  perform  a special similarity transformation
\begin{eqnarray}
L(k,I) \;\; \Longrightarrow  \;\;  L_{isotr}(k,I) = S L(k,I)S^{-1}
\; , \nonumber
\\
L(I,m) \;\; \Longrightarrow  \;\;  L_{isotr}(I,m) =S L(I,m)S^{-1}
\; , \nonumber
\\
L(k,m) \Longrightarrow U(k,m) =  L(k,m)_{isotr} = SL(k,m)S^{-1} \;
, \label{15.2a}
\end{eqnarray}

\noindent  where
\begin{eqnarray}
S = {1 \over \sqrt{2}} \left | \begin{array}{rrrr}
1 & 0  & 0  &  1 \\
1 & 0  & 0  &  - \\
0 & 1  & -i  & 0 \\
0 & 1  &  i  & 0 \\
\end{array} \right | ,
\qquad S ^{-1} = {1 \over \sqrt{2}} \left | \begin{array}{rrrr}
1 & 1  & 0  &  0 \\
0 & 0  & 1  &  1 \\
0 & 0  & i  & -i \\
1 & -1  &   0  & 0 \end{array} \right | ,
\nonumber
\end{eqnarray}

\noindent in  the isotropic basis we will obtain
\begin{eqnarray}
L_{isotr}(k,I) = \left | \begin{array}{cccc}
(k_{0}-k_{3})      &    0     &   -(k_{1} +ik_{2})  &    0  \\
0    & (k_{0}+k_{3})  &   0    &  -(k_{1} -ik_{2})   \\
-(k_{1}-ik_{2})  & 0  & (k_{0} +k_{3}) & 0 \\
0 & -(k_{1}+ik_{2})   & 0  & (k_{0} -k_{3})
\end{array} \right | \; ,
\nonumber
\\
L_{isotr}(I,m) = \left | \begin{array}{cccc}
(m_{0}-m_{3})      &    0     &  0 &   -(m_{1} -im_{2})    \\
0    & (m_{0}+m_{3})  &  -(m_{1} +im_{2}) & 0    \\
0 & -(m_{1}-im_{2})  &  (m_{0} -m_{3}) & 0 \\
-(m_{1}+im_{2})   & 0  & 0 & (m_{0} +m_{3})
\end{array} \right | \; .
\nonumber
\end{eqnarray}

\noindent
It is convenient to employ the other notation
referring to  initial  spinor matrices:
\begin{eqnarray}
B(k) =  \left |  \begin{array}{cc}  a & d \\
c & b \end{array} \right | \; , \qquad B(\bar{m}) =
\left |
\begin{array}{cc} B & -C \\ -D & A
\end{array} \right |  \; ,
\nonumber
\end{eqnarray}

\noindent  then  the previous relations   will take the form
\begin{eqnarray}
L_{isotr}(k,I) = \left | \begin{array}{rrrr}
b     &    0     &   -c  &    0  \\
0    & a  &   0    &  -d   \\
-d  & 0  & a & 0 \\
0 & -c   & 0  & b
\end{array} \right | \; , \qquad
L_{isotr}(I,m) = \left | \begin{array}{rrrr}
B     &    0     &   0  &    -C  \\
0    & A  &   -D  &   0  \\
0  & -C  & B & 0 \\
-D & 0   & 0  & A
\end{array} \right | \; .
\label{15.5a}
\end{eqnarray}

\noindent
The Lorentz  complex vector matrix  $L_{isotr} (k,m) $ in isotropic  basis looks as
\begin{eqnarray}
L_{isotr} (k,m) = \left | \begin{array}{rrrr}
bB  & cC & -cB & -bC \\
dD & aA  & -aD & -dA \\
-dB & -aC & aB & dC \\
-bD & - cA & cD & bA
\end{array} \right | \in SO(3.1.C) \; .
\label{15.5b}
\end{eqnarray}

\noindent
Restriction to real Lorentz  group is achieved by a formal
substitution:
\begin{eqnarray}
A \longrightarrow a^{*},  \qquad B \longrightarrow b^{*},  \qquad
C \longrightarrow c^{*},  \qquad D \longrightarrow d^{*},  \qquad
\nonumber
\\
L_{isotr} (k,k^{*}) = \left | \begin{array}{rrrr}
bb^{*}  & cc^{*} & -cb^{*} & -bc^{*} \\
dd^{*} & aa^{*}  & -ad^{*} & -da^{*} \\
-db^{*} & -ac^{*} & ab^{*} & dc^{*} \\
-bd^{*} & - ca^{*} & cd^{*} & ba^{*}
\end{array} \right | \in SO(3.1.R) \; .
\label{15.5b}
\end{eqnarray}

In contrast to the usual (non isotropic) basis, here  the Lorentz
matrix consists of 16 elements each of  these is constructed as a
single multiple of two quantities. In the  case of ordinary basis
each matrix element  represent  a sum of  four such combinations.
It is  this
 feature that  makes  the isotropic basis so helpful in applications.

\section{The  covering for  $SO(3.1.C)$  and intrinsic fermion parity}

\hspace{5mm}
The problem of intrinsic parity for a fermion has a long history
(see Introduction),   the problem seems to be unsolved till now.
  The main difficulty consists in double valuedness of spinors considered in the frames
  of   orthogonal groups.
  In our opinion, the problem should be studied
  on the basis of the representation theory for spinor covering of the orthogonal
  groups.\footnote{In essence,  a strong form of this changing orthogonal groups to their covering groups
and  its  representations throughout,
 and when we  are going to work in the same manner  at describing
 the space-time structure itself, is the known Penrose-Rindler spinor  approach [163-167].}.
Below we propose the way to solve this problem through extension
 of the covering  for complex Lorentz group  $SO(3.1.C)$
by adding two discrete 4-spinor operations,
$P= i\gamma^{0}$  and  $T = \gamma^{0}\gamma^{5}$ can be quite easily  solved\footnote{The choice of phase factors
at these two matrices is in accordance with Racah choice [39]. In any of Majorana bases these
discrete operations are given by real matrices.}.
 One may search pure spinor representations of the covering group
 $G_{cover} = \{ G \oplus P \oplus T \}$ in the form
 \begin{eqnarray}
T(g) = s(g)\; g\; , \;\;\; g \in  G_{cover} \; ,\;\;
 s(g_{1}) \; s(g_{2}) =  s(g_{1} \;g_{2})    \; ,
\label{25}
\end{eqnarray}

\noindent  where  $s(g)$ is a numerical function on the group.
There exist four different solutions  $s_{i}(g)$:
\begin{eqnarray}
\left.  \begin{array}{ccccc}
G_{cover} & s_{1}(g) =    &    s_{2}(g) = &   s_{3}(g)  =     &    s_{4}(g) =   \\  [2mm]
G(k,m)      &     + 1       &    + 1   &  + 1   &    + 1 \;         \\
P                     &     + 1       &    - 1   &  + 1   &    - 1  \;        \\
T                    &     + 1       &    - 1   &  - 1   &    + 1 \; .
\end{array} \right.
\label{26}
\end{eqnarray}

\noindent Correspondingly we arrive at four type of representations $S_{i}(g)$
of the group   группы $G_{cover}$. With the use of the relations
$$
F = \mbox{const}        \left | \begin{array}{cc}
                           - I &  0   \\  0 & + I        \end{array} \right | \;, \qquad
F \; G(k,m) \; F^{-1} = G(k, m) \; ,
$$
$$
F \; P\;  F^{-1} = - P \; , \qquad
 F \;  T'\; F^{-1} = - T\; ,
$$

\noindent one can conclude that representation  $S_{2}(g)$  is equivalent to  $S_{1}(g)$,
and  $S_{4}(g)$  is equivalent to  $S_{3}(g)$:
\begin{eqnarray}
S_{2}(g) = F \; S_{1}(g) \;F^{-1} \;, \qquad S_{4}(g) = F \; S_{3}(g)\;
F^{-1} \;.
\label{27}
\end{eqnarray}

Thus, there exist only two different types of the pure spinor representations
of the covering group
\begin{eqnarray}
S_{1}(g)  \sim S_{2}(g)  \;\;  ,  \;\; S_{3}(g)  \sim S_{4}(g) \; .
\label{28}
\end{eqnarray}

This doubling can be connected with intrinsic
 space-time parity for a fermion. Special  attention should be given to the following:
 no separate $P$-parity or $T$-parity for a fermion exist in  the  group-theoretical sense,
 instead the group theory  leads to a unified intrinsic characteristic as  a fermion's  parity.

Manifestation of these two discrete spinor operations in tensor representation
one can examine by looking at 2-rank 4-spinor:
\begin{eqnarray}
\Psi' = G \Psi\; , \qquad  \Phi ' = G \Phi\; , \qquad
\Psi \otimes \Phi = U \; ,  \qquad U' = G U \tilde{G} \; ,
\nonumber
\\
U = [\; \varphi \; I + i \tilde{\varphi} \; \gamma^{5} + i \varphi_{l}\; \gamma^{l} +
 \tilde{\varphi}_{l}\; \gamma^{l}
\gamma^{5} + \varphi_{mn} \; \sigma_{mn} \; ]\; \gamma^{0}\gamma^{2} \; ;
\label{23}
\end{eqnarray}

\noindent
the symbol of tilde over $G$   stands for the matrix transposition.
$\gamma^{0}\gamma^{2}$  is a 4-spinor
metric matrix. The transformations properties of
irreducible tensor components of 2-rank 4-spinor are given by
\begin{eqnarray}
\left. \begin{array}{lll}
\underline{G: } \qquad  & \varphi ' = \varphi \; , \qquad   &
\tilde{\varphi} ' = \tilde{\varphi } \; ,\\[2mm]
& \varphi_{k}' = L_{k}^{\;\; l} \varphi_{l} \; , \qquad &
\tilde{\varphi}_{k}' = L_{k}^{\;\; l} \tilde{\varphi}_{l} \; ,\\[2mm]
& \varphi_{kl}' = L_{k}^{\;\; m}  L_{l}^{\;\; n}  \;\varphi_{mn} \; ,  &
\end{array} \right.
\nonumber
\\
\left. \begin{array}{lll}
\underline{P: } \qquad  & \varphi ' = +\varphi \; , \qquad   &
\tilde{\varphi} ' = -\;\tilde{\varphi } \; ,\\[2mm]
& \varphi_{k}' = + L_{k}^{(P) l} \varphi_{l} \; , \qquad &
\tilde{\varphi}_{k}' = - L_{k}^{(P) l} \tilde{\varphi}_{l} \; ,\\[2mm]
& \varphi_{kl}' = + L_{k}^{(P) m}  L_{l}^{(P) n}  \;\varphi_{mn} \; ,  &
\end{array} \right.
\nonumber
\\
\left. \begin{array}{lll}
\underline{T: } \qquad  & \varphi ' = -\varphi \; , \qquad   &
\tilde{\varphi} ' = +\;\tilde{\varphi } \; ,\\[2mm]
& \varphi_{k}' = -L_{k}^{(T) l} \varphi_{l} \; , \qquad &
\tilde{\varphi}_{k}' = + L_{k}^{(T) l} \tilde{\varphi}_{l} \; ,\\[2mm]
& \varphi_{kl}' =  -L_{k}^{(T) m}  L_{l}^{(T) n}  \;\varphi_{mn} \; ,  &
\end{array} \right.
\label{24}
\end{eqnarray}

where
$$
L_{c}^{(P)a} = \bar{\delta}_{c}^{a} =
 \left | \begin{array}{cccc}
1 & 0 & 0 & 0 \\
0 & -1 & 0 & 0 \\
0 & 0 & -1 & 0 \\
0 & 0 & 0 & -1
\end{array} \right |\; , \qquad
L_{c}^{(T)a} = - \; \bar{\delta}_{c}^{a} =
 \left | \begin{array}{cccc}
-1 & 0 & 0 & 0 \\
0 & 1 & 0 & 0 \\
0 & 0 & 1 & 0 \\
0 & 0 & 0 & 1
\end{array} \right |\; .
$$

In order to construct tensors with different properties under discrete operations
one should take into account the possibility  to distinguish  4-spinor with respect
 to intrinsic parity.

The 2-rank 4-spinor of the type $
U_{13} = \Psi_{1} \otimes \Phi _{3} \sim  \Psi_{3} \otimes \Phi _{1}
$
contains irreducible tensors with the following properties:
\begin{eqnarray}
\left. \begin{array}{lll}
\underline{G: } \qquad  & \varphi ' = \varphi \; , \qquad   &
\tilde{\varphi} ' = \tilde{\varphi } \; ,\\[2mm]
& \varphi_{k}' = L_{k}^{\;\; l} \varphi_{l} \; , \qquad &
\tilde{\varphi}_{k}' = L_{k}^{\;\; l} \tilde{\varphi}_{l} \; ,\\[2mm]
& \varphi_{kl}' = L_{k}^{\;\; m}  L_{l}^{\;\; n}  \;\varphi_{mn} \; ,  &
\end{array} \right.
\nonumber
\\
\left. \begin{array}{lll}
\underline{P: } \qquad  & \varphi ' = +\varphi \; , \qquad   &
\tilde{\varphi} ' = -\;\tilde{\varphi } \; ,\\[2mm]
& \varphi_{k}' = + L_{k}^{(P) l} \varphi_{l} \; , \qquad &
\tilde{\varphi}_{k}' = - L_{k}^{(P) l} \tilde{\varphi}_{l} \; ,\\[2mm]
& \varphi_{kl}' = + L_{k}^{(P) m}  L_{l}^{(P) n}  \;\varphi_{mn} \; ,  &
\end{array} \right.\nonumber
\\
\left. \begin{array}{lll}
\underline{T: } \qquad  & \varphi ' = +\varphi \; , \qquad   &
\tilde{\varphi} ' = -\;\tilde{\varphi } \; ,\\[2mm]
& \varphi_{k}' = +L_{k}^{(T) l} \varphi_{l} \; , \qquad &
\tilde{\varphi}_{k}' = - L_{k}^{(T) l} \tilde{\varphi}_{l} \; ,\\[2mm]
& \varphi_{kl}' =  +L_{k}^{(T) m}  L_{l}^{(T) n}  \;\varphi_{mn} \; ,  &
\end{array} \right.
\label{29}
\end{eqnarray}

\noindent
Thus all four type of scalars and vectors,  $(++),
(--),(+-),(-+)$ have been constructed.

Now we consider the problem of linear representations of the spinor groups that
are supposedly cover the partly extended Lorentz groups
$L_{+-}^{\uparrow}$    and  $L_{+}^{\uparrow\downarrow}$  (improper orthochronous and
 proper non-orthochronous respectively).
Such covering of partly extended Lorentz group  can be constructed by adding only
 one matrix discrete operation: $P$ or $T$.
The result obtained for simplest  representations of these groups
 is as follows:
all  above representations $S_{i}(g)$    at confining them to sub-groups
$\{ G \oplus  P \} $   or  $\{ G \oplus T\} $   lead to  representations changing into each other
by a similarity
transformation. In other words,  in fact  there exists only one representation of these
partly extended spinor groups. This may be understood as  impossibility to determine any group-theoretical
parity concept ($P$ or $T$) within the limits of partly extended spinor
 groups\footnote{This is in evident contradiction
with the analysis of fermion parity given by Yang and Tiomno [54]  and
 many others in the frames of
orthogonal groups theory.}

\section{On the structure of Majorana bases}

\hspace{5mm}
Additionally, several points in connection to the  Majorana bases
in the context
of  the real Lorentz group  should be added. In particular, any Lorentz transformation
in bispinor space is resolved in terms
of the Dirac matrices [101]:
\begin{eqnarray}
S(k,\; k^{*}) =   {1 \over 2} (k_{0}  +  k^{*}_{0})  -
{1\over 2} ( k_{0} -  k^{*}_{0})  \gamma ^{5}  +
k_{1}  (\sigma ^{01}  +  i \sigma ^{23})  +
k^{*}_{1} (\sigma ^{01} -  i\sigma ^{23})   +
\nonumber
\\
k_{2} (\sigma ^{02}  +  i\sigma ^{31})   +
k^{*}_{2} (\sigma ^{02}  -  i\sigma^{31})  +
 k_{3}  (\sigma ^{03} +  i\sigma ^{12})   +  k^{*}_{3} (\sigma
^{03}  -  i\sigma ^{12})   \; .
\label{23}
\end{eqnarray}

 \noindent
Because in Majorana basis $
 ( i \gamma ^{5}_{M} )^{*} = i \gamma ^{5}_{M}  \; ,  \;  ( \sigma ^{ab}_{M}  )^{*}  =
+ \; \sigma ^{ab}_{M} \;,
$
the Lorentzian 4-spinor matrix  $S(k, \;k^{*})$  is real.

Now we are to describe all possible Majorana's bases.
To this end we  should find  all transformations $A$ in 4-spinor  space
that change all the Dirac matrices  to the  imaginary forme
\begin{eqnarray}
\Psi_{M} (x) = A \;  \Psi (x) \; ,
\qquad
\gamma ^{a}_{M} = A \; \gamma ^{a} \; A^{-1} \; , \qquad ( \gamma
^{a}_{M})^{*} = - \gamma _{M} ^{a}   \; ;
\label{1.18.1}
\end{eqnarray}

\noindent here  $\gamma ^{a}$  stand for the Dirac matrices in spinor representation.
One  should note  that the problem must have  a many of solutions. Indeed,
if  $A$ satisfies eq. (\ref{1.18.1}) then any  matrix of the form
$A' =  e^{i\alpha } \; R\;  A $, where  $R$  is  real,  will satisfy eq.
(\ref{1.18.1}) as well:
\begin{eqnarray}
\mbox{if} \hspace{50mm} (A \; \gamma^{a}\;A^{-1})^{*} = - (A \; \gamma^{a}\;A^{-1})   \; ,
\hspace{30mm}
\nonumber
\\
\mbox{then}\qquad\qquad \left [\; (e^{i\alpha}\; R\; A )\; \gamma^{a} \; (e^{i\alpha} \; R
\; A )^{-1} \; \right ]^{*} = - \left [\; (e^{i\alpha}\; R \; A
)\; \gamma^{a} \; (e^{i\alpha}\; R \; A )^{-1} \; \right ]^{*} \;
.
\nonumber
\end{eqnarray}

Equation  (\ref{1.18.1}) can be written as
\begin{eqnarray}
A^{*}\; (\gamma^{a})^{*} \;(A^{*})^{-1} = - A\; \gamma^{a} \;
A^{-1}  \; ,\qquad \mbox{or} \qquad
(A^{-1} \; A^{*})  \; (\gamma^{a})^{*} \;  (A^{-1} \; A^{*})^{-1}
= - \; \gamma^{a} \; .
\nonumber
\end{eqnarray}

\noindent With the  use of notation
$
A^{-1} \; A^{*}   = U $,
 the latter reads
\begin{eqnarray}
U \;  (\gamma^{a})^{*} \;  U^{-1} = - \;\gamma^{a} \; .
\label{1.18.2b}
\end{eqnarray}

\noindent In spinor representation four  identities hold
\begin{eqnarray}
(\gamma^{0})^{*} = + \gamma^{0} \; ,  \;\; (\gamma^{1})^{*} = +
\gamma^{1} \; ,  \;\; (\gamma^{2})^{*} = - \gamma^{2} \; ,  \;\;
(\gamma^{3})^{*} = + \gamma^{3} \; ,
\nonumber
\end{eqnarray}

\noindent therefore, solution of  eq. (\ref{1.18.2b}) looks as
\begin{eqnarray}
U = const \; \gamma^{2} \; , \qquad \mbox{det} \;  (\gamma^{2})  = +1\; .
\label{1.18.2c}
\end{eqnarray}

\noindent Because  $\mbox{det} \; U =  (det \; A)^{*} /
(det \; A)$  ,
one  must conclude that $const$ is equal to a phase factor  $e^{i\alpha}$.
Thus, the problem is reduced to
\begin{eqnarray}
A^{-1} \; A^{*} = e^{i \alpha}\; \gamma^{2} \; ,\;\; \; \mbox{or} \;\;\;
A^{*} = e^{i \alpha} \; A \; \gamma^{2} \; .
\label{1.18.3b}
\end{eqnarray}

Any 4-dimensional matrix can be decomposed into sixteen Dirac elementary matrices
\begin{eqnarray}
A = \left [ \; (M_{0} \gamma^{2} + m_{0})  \; + \; \gamma^{1} \;
(N_{0} \gamma^{2} + n_{0}) \; \right ]  \; +
\nonumber
\\
+ \; \gamma^{5} \; \left [ \; (M_{1} \gamma^{2} + m_{1})  \; + \;
\gamma^{1}\; (N_{1} \gamma^{2} + n_{1}) \; \right ] \; +
\nonumber
\\
+\; \gamma^{0} \; \left [ \; (M_{2} \gamma^{2} + m_{2})  \; + \;
\gamma^{1} \; (N_{2} \gamma^{2} + n_{2}) \;  \right ]  \; +
\nonumber
\\
+ \; \gamma^{5} \gamma^{0} \; \left [ \; (M_{3} \gamma^{2} +
m_{3})  \; + \; \gamma^{1}\; (N_{3} \gamma^{2} + n_{3}) \; \right
] \; .
\label{1.18.4a}
\end{eqnarray}

\noindent With the notation
\begin{eqnarray}
\Gamma_{0} = I \; , \;\; \Gamma_{1} = \gamma^{5} \; , \;\;
\Gamma_{2} = \gamma^{0}\; , \;\; \Gamma_{3} =\gamma^{5}
\gamma^{0}\; ,
\label{1.18.4b}
\end{eqnarray}

\noindent  eq.  (\ref{1.18.4a}) is written in the abridged form
\begin{eqnarray}
A = \Gamma_{i} \; \left [ \; (M_{i} \; \gamma^{2} + m_{i})  \; +
\; \gamma^{1}\; (N_{i} \;  \gamma^{2} + n_{i}) \; \right ] \; .
\label{1.18.4c}
\end{eqnarray}

\noindent Now, let us  substitute  (\ref{1.18.4c}) into eq. (\ref{1.18.3b}):
\begin{eqnarray}
\Gamma_{i} \; \left [ \;  (-M_{i}^{*} \gamma^{2} + m_{i}^{*}) \; +
\; \gamma^{1}\; (-N_{i}^{*} \gamma^{2} + n_{i}^{*}) \; \right ] =
\nonumber
\\
= e^{i \alpha } \Gamma_{i} \; \left [ \; (m_{i} \gamma^{2} -
M_{i})  \; + \; \gamma^{1}\; (n_{i} \gamma^{2} - N_{i}) \; \right
] \; ;
\nonumber
\end{eqnarray}

\noindent further it follow equations for unknown parameters:
\begin{eqnarray}
M_{i}= - e^{-i\alpha}\;  m_{i}^{*} \; , \qquad N_{i}= -
e^{-i\alpha}  \; n_{i}^{*} \; .
\nonumber
\end{eqnarray}

\noindent Therefore, expression for a  matrix  $A$,
relating spinor basis with any one Majorana's is given by
\footnote{All the matrices in the right-hand side are referred to
 spinor representation.}
\begin{eqnarray}
A =  \Gamma_{i} \; \left [  \; (m_{i} - e^{-i\alpha}\;  m_{i}^{*}
\;\gamma^{2})\; +\; \gamma_{1} \; (n_{i} - e^{-i\alpha} \;
n_{i}^{*} \;\gamma^{2}) \; \right ] \; ,
\nonumber
\\
\Psi_{M} (x) = A(m_{i}, n_{i}, e^{i\alpha} ) \;  \Psi (x) \; . \hspace{20mm}
\label{1.18.5}
\end{eqnarray}

\noindent Evidently, 17 arbitrary real parameters enter these formulas;
 there exists one additional restriction,
 $\mbox{det} \; A \neq 0$.
Simplest  of the Majorana bases  can be  taken as follows
(only different from zero parameters are specified):
\begin{eqnarray}
\underline{m_{0} = 1 /  \sqrt{2},  \qquad e^{i\alpha} = +1 \; } , \qquad
A= {1 - \gamma^{2} \over \sqrt{2}} \; , \qquad A^{-1}= {1 +
\gamma^{2} \over \sqrt{2}} \;,
\nonumber
\\
\gamma^{0}_{M} = +\gamma^{0} \gamma^{2} , \qquad \gamma^{1}_{M} =
+ \gamma^{1} \gamma^{2} , \qquad \gamma^{2}_{M} =  \gamma^{2} ,
\qquad \gamma^{3}_{M} = + \gamma^{3} \gamma^{2} .
\end{eqnarray}

There exist 16  such possibilities, they can be listed in the
 two tables:

$ \underline{e^{i\alpha} = +1} $
\begin{eqnarray}
\left. \begin{array}{lcccccccc} 1/\sqrt{2}=   &
\qquad m_{0}& m_{1} & m_{2} & m_{3}    & \qquad  n_{0} &  n_{1} & n_{2} & n_{3} \\[3mm]
\gamma^{0}_{M} = \gamma^{0} \gamma^{2} \times
                                       & \qquad  +1    &  +1    &  -1   &  -1
                                       & \qquad  +1    &  +1    &  -1   &  -1 \\[2mm]
\gamma^{1}_{M} = \gamma^{1} \gamma^{2} \times
                                       & \qquad  +1    &  +1    &  +1   &   +1
                                       & \qquad  -1    &  -1    &  -1   &   -1 \\[2mm]
\gamma^{2}_{M} = \gamma^{2} \times
                                       & \qquad  +1    &  -1    &  -1   &   +1
                                       & \qquad  -1    &  +1    &  +1   & -1 \\[2mm]
\gamma^{3}_{M} = \gamma^{3} \gamma^{2} \times
                                       & \qquad  +1    &  +1    &  +1   & +1
                                       & \qquad  +1    &  -1    &  +1   & +1
\end{array} \right.
\label{1.18.23a}
\end{eqnarray}

$ \underline{e^{i\alpha} = -1} $ \begin{eqnarray}
\left. \begin{array}{lcccccccc} 1/\sqrt{2}=   &
\qquad m_{0}& m_{1} & m_{2} & m_{3}    & \qquad  n_{0} &  n_{1} & n_{2} & n_{3} \\[3mm]
\gamma^{0}_{M} = \gamma^{0} \gamma^{2} \times
                                       & \qquad  -1    &  -1    &  +1   &  +1
                                       & \qquad  -1    &  -1    &  +1   &  +1 \\[2mm]
\gamma^{1}_{M} = \gamma^{1} \gamma^{2} \times
                                       & \qquad  -1    &  -1    &  -1   &   -1
                                       & \qquad  +1    &  +1    &  +1   &   +1 \\[2mm]
\gamma^{2}_{M} = \gamma^{2} \times
                                       & \qquad  +1    &  -1    &  -1   &   +1
                                       & \qquad  -1    &  +1    &  +1   & -1 \\[2mm]
\gamma^{3}_{M} = \gamma^{3} \gamma^{2} \times
                                       & \qquad  -1    &  -1    &  -1   & -1
                                       & \qquad  -1    &  +1    &  -1   & -1
\end{array} \right.
\label{1.18.24a}
\end{eqnarray}

\noindent
These  Majorana bases are similar to each other:
 $\gamma^{2}_{M} =  \pm  \gamma^{2} $ , the remaining three  Dirac matrices
are multiplied by $\pm \gamma^{2}$.

\section{Conclusions }

A  unifying overview of the ways to parameterize the linear group
$GL(4.C)$ and its subgroups  is given. As parameters for this
group   there are taken 16 coefficients $G
=G(A,B,A_{k},B_{k},F_{kl})$ in resolving matrix  $G\in GL(4.C)$ in
terms of 16 basic elements of the Dirac matrix algebra. The
multiplication rules $G'G$ are formulated in the form of a
bilinear function of two sets of 16 variables. The detailed
investigation is restricted to  6-parameter case $G(A,B,F_{kl})$,
which provides  us with spinor  covering for   the complex
orthogonal  group  $SO(3.1.C)$. Lorentz  matrices $L$ in complex
space  are determined through the formula $G\gamma^{a}G^{-1}=
\gamma^{c} L_{c}^{\;\;a}, L = L(A,B,F_{kl})$. Restrictions to
parameters corresponding to spinor coverings of sub-groups $
SO(3.1.R),SO(2.2.R),$ $SO(4.R), SO(3.C), SO(3.R), SO(2.1.R) $
 are formulated explicitly.
 The use of the
Newmann-Penrose formalism  and applying quaternion techniques in
the theory of complex Lorentz group are  discussed.
 The complex Euler's  angles parametrization for  complex Lorentz group
 is elaborated as underlaid by  the orthogonal  complex cylindrical coordinates in the
complex  extension for Riemannian 3-space of constant curvature.
The Majorana bases are studied in detail. In an explicit form
17-parametric formula referring spinor (Weyl's) basis to any
Majorana's one has been established: $\Psi_{M} (x) = A(m_{i},
n_{i}, e^{i\alpha} ) \;  \Psi (x)$. The most simple choices of
($m_{i}, n_{i}, e^{i\alpha}$) associated with widely used Majorana
representations  are given.

\noindent The problem of extending the group $SO(3.1.C)$  by two
additional discrete spinor operations,   $P$ and    $T$
reflections:
 $G_{cover}= \{G \oplus P \oplus T \}$ is solved. It is shown that the extended
  covering group   $G_{cover}$ has only two types of irreducible fundamental 4-spinor
  representations, which  can serve the group-theoretical base for
  the concept of intrinsic parity of 4-spinor field. Resolving the 2-rank spinor
$\Psi \otimes \Phi$ into a set of irreducible components under the
covering group $G_{cover}$ is given;  four types  of scalars and
4-vectors are constructed.
 The problem of extending the group looks the same for  $  O(3.1.C)$, $SO(3.1.R),SO(2.2),SO(4.R) $.
 Connections
between  Einstein-Mayer  study on semi-vectors and Fedorov's
treatment of the Lorentz group theory are stated in detail.

\vspace{3mm}

The work is dedicated  to the 95-th anniversary of birthday of
Academican F.I. Fedorov.


\end{document}